\documentclass[conference]{IEEEtran}
\IEEEoverridecommandlockouts
\usepackage{cite}
\usepackage{amsmath,amssymb,amsfonts}
\usepackage{algorithmic}
\usepackage{graphicx}
\usepackage{textcomp}
\usepackage[dvipsnames]{xcolor}
\usepackage{ragged2e}

\usepackage[utf8]{inputenc} 
\usepackage[T1]{fontenc}    
\usepackage{hyperref}       
\usepackage{url}            
\usepackage{booktabs}       
\usepackage{amsfonts}       
\usepackage{nicefrac}       
\usepackage{microtype}      
\usepackage{graphicx}
\usepackage{float}
\usepackage{amsmath}
\usepackage{wrapfig}
\usepackage{bm}
\usepackage{subcaption}
\usepackage[numbers]{natbib}
\usepackage{enumitem}

\usepackage{subcaption}
\usepackage{array}
\usepackage{multirow}

\usepackage{titlesec}
\def\BibTeX{{\rm B\kern-.05em{\sc i\kern-.025em b}\kern-.08em
    T\kern-.1667em\lower.7ex\hbox{E}\kern-.125emX}}
\begin{document}

\title{Coupling Deep Imputation with Multitask Learning for Downstream Tasks on Omics Data}


 \author{\IEEEauthorblockN{Sophie Peacock}
 \IEEEauthorblockA{\textit{AstraZeneca}\\
 Cambridge, UK \\
 sophie.peacock@astrazeneca.com}
 \and
 \IEEEauthorblockN{Etai Jacob}
 \IEEEauthorblockA{\textit{AstraZeneca}\\
 Waltham, USA \\
 etai.jacob@astrazeneca.com}
 \and
 \IEEEauthorblockN{Nikolay Burlutskiy}
 \IEEEauthorblockA{\textit{AstraZeneca}\\
 Cambridge, UK \\
 nikolay.burlutskiy@astrazeneca.com}
 }

\maketitle

\begin{abstract}
    Omics data such as RNA gene expression, methylation and micro RNA expression are valuable sources of information for various clinical predictive tasks. For example, predicting survival outcomes, response to drugs, cancer histology type and other patients related information is possible using not only clinical data but molecular data as well. Moreover, using these data sources together, for example in multitask learning, might boost the performance. However, in practice, there are many missing data points which leads to significantly lower patient numbers when analysing full cases, which in our setting refers to all modalities being present. 
    
    In this paper we investigate how imputing data with missing values using deep learning coupled with multitask learning can help to reach state-of-the-art performance results using combined omics modalities - RNA, micro RNA and methylation. We propose a generalised deep imputation method to impute values where a patient has data for one modality missing. Interestingly, deep imputation by itself outperforms multitask learning in classification and regression tasks across most combinations of modalities. In contrast, when using all available modalities for survival prediction we observe that multitask learning by itself significantly outperforms deep imputation (adjusted p-value of 0.03). Thus, both approaches are complementary when optimising performance for downstream predictive tasks.
\end{abstract}


\section{Introduction and Related Work}\label{relatedWork}

The advancements in high-throughput molecular technologies in the last decade have made it possible to measure concurrently genomics, transcriptomics, epigenetics, proteomics and other omics modalities to obtain a more complete picture of cancer patients. The breadth of modalities in this space have stimulated much active research in combining multi-omics data for predicting end points such as survival, patient outcomes, and clinical features \citep{zhang2021integrating, baek2020prediction,ma2019affinitynet,chaudhary2018deep}. Important to mention are studies which included clinical features such as age and cancer type for survival predictions \citep{zhao2021pan,brandt2015age}. While these features have a relatively strong prognostic power for survival predictions, they can mask the value of -omics data which enable researchers to unravel the biology behind the disease. In this work, we sought to maximise the value of omics data by implicitly including representative clinical features in our model, in a manner which minimises their masking effect on omics data.

Most real-world and clinical study data sets have missing values, within or across modalities, which significantly limits analysis and motivates the search for new approaches to deal with missing values. For example, in many clinical trials and real world practices, only a subset of genes are included in a panel that is used to measure the RNA expression levels or DNA mutations from patients derived blood samples or a biopsy from the tumour. This question is especially critical when working with multi-modal data where one has several modalities measured for a single patient. An immediate solution is to simply remove patients with missing values, which could result in a substantial decrease of sample size in the analysis. Another approach is to perform some sort of naive imputation, for example, mean imputation \citep{donders2006gentle}. There have been few studies which developed imputation methods for reconstructing one modality from another \citep{dong2019tobmi,hu2019statistical,zhong2019predicting}. Deep learning methods allow one to reconstruct complex non-linear relationships between modalities as well as missing values within a modality. An important work by \citet{zhou2020imputing} demonstrated how a deep learning autoencoder based architecture enables the reconstruction of RNAseq information from methylation data. Inspired by this work, we generalise this method to multiple modalities, including methylation and micro-RNA (miRNA). By applying our method to TCGA data we demonstrate its advantage in the multi-modal integration setting.

The Cancer Genome Atlas (TCGA) is a rich publicly available data set that provides researchers with harmonised multi-modal data such as gene expression, clinical outcomes, DNA methylation and other modalities \citep{tcga}. The data set includes approximately 11,000 patients with data covering a 12 year period and comprises 33 cancer types.

Multiple studies have suggested ways of combining different modalities with the aim of boosting the performance of downstream tasks \citep{hyun2019machine,yala2019deep,reda2018deep}. Such methods include early, intermediate and late fusion approaches which relates to the step at which the data modalities are combined. In early fusion, the data modalities are concatenated and the single data source propagated through a ML pipeline. With late fusion the data sources are first passed through independent ML pipelines and the resulting lower dimension sources are merged and passed through a shallow inference model. In many cases, deep learning approaches demonstrate superiority over classic, lower complexity machine learning methods especially when the data sets are large enough\citep{ahmad2019deep}. Nevertheless, methods for optimal combination of multiple modalities are under active research \citep{eicher2020metabolomics}. Some modalities have demonstrated synergies and others redundancies. Also, some modalities are more noisy than others by design \citep{kim2015nature}.

A common paradigm of overcoming the problem of high dimensional data in  ML is to feed models a compressed version of the original space following dimensionality reduction. This procedure is a fundamental part of the pipeline and can result in a detrimental impact on downstream performance if the projected latent space is an over-fitted or under-fitted representation of the original space. To ensure the latent space is generalisable enough for multiple tasks and is a good representation of the original data, multitask learning is a popular approach. Multitask learning is an active field of machine learning and is based on the simultaneous training of several tasks. In such settings, the trained models yield a more generalisable latent space which facilitates better performance of downstream tasks. For example, \citet{zhang2021omiembed} demonstrated that applying multitask learning to TCGA multi-omics data led to better performance compared to training on independent modalities or training for a single task. However, the missing values in the analysed data were imputed using mean imputation which is an oversimplification. In this paper, we apply imputation using a deep learning method which provides an improved reconstruction of the input data.

We demonstrate an approach for accurately and efficiently solving multiple downstream tasks by maximizing the use of available data in the presence of missing values. We first generalised the work of \citep{zhou2020imputing} by imputing data for three different modalities, RNAseq $\rightarrow$ methylation, RNAseq $\rightarrow$ miRNA and methylation $\rightarrow$ RNAseq. We imputed missing values on TCGA data, significantly increasing the number of patients with complete records. We imputed RNAseq data for 371 patients, methylation data for 1,358 patients and miRNA data for 173 patients, leading to a 19\% increase in overall patient numbers. We then investigated how a) deep imputation, b) multitask learning c) deep imputation coupled with multitask learning compared to single modalities or their combinations for three different downstream tasks. We compared performance using both the original data set and the larger imputed data set. Finally, we outline a recommendation for training multi-modal models to achieve the best performance on downstream clinical end point tasks.

\section{Methodology}\label{methods}

\subsection*{Deep Imputation}

\begin{figure}[tb]
\begin{center}
\includegraphics[width=0.6\linewidth]{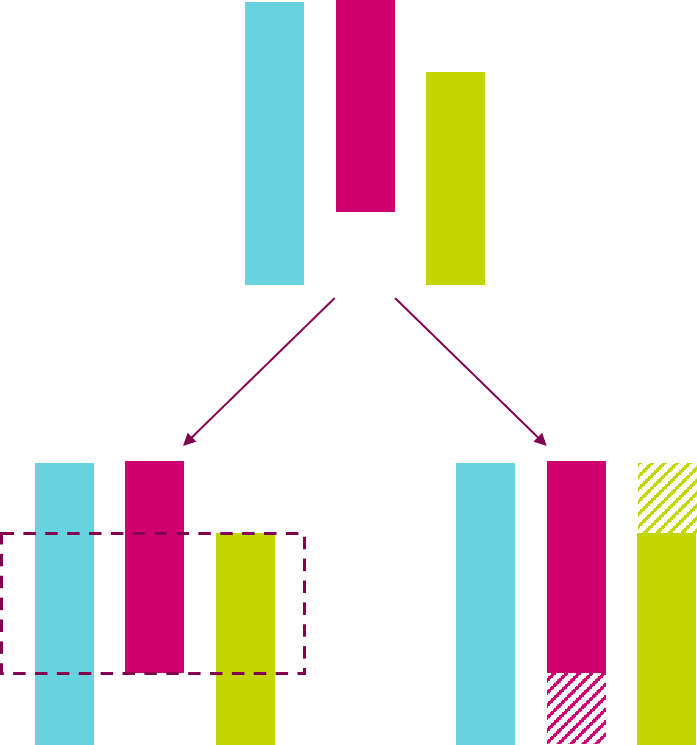}
\caption{Benefits of imputation. In the top image two out of the three modalities are missing for some patients. With no imputation, we can only include patients in the dashed box who have data for all three modalities (left image). This can be resolved by using imputation (right image).}
\label{fig:imputation_diagram}
\end{center}
\end{figure}

Mean imputation is a simple and popular method for filling in missing values for machine learning tasks. This technique introduces systematic bias due to the oversimplification of the mean operation and does not take advantage of multivariate information found in the data set. Other imputation techniques which take into consideration the multivariate nature of the data, such as k-nearest neighbour and singular value decomposition, achieve good results when data is missing at random positions. However, these methods under-perform when an entire set of values or modality is missing for an entry, for example, at a patient level \citep{troyanskaya2001missing}. The need for a better method of imputing missing values is even more pronounced in human derived data acquisition cases, where a segment of patients have whole modalities missing from their records. Instead of having to remove these patients from the cohort it would be beneficial to be able to impute the values and avoid reducing the size of the data set. A demonstration of the potential effect of such an approach on data set size is shown in Figure \ref{fig:imputation_diagram}. We therefore carried out a set of experiments to discover whether such methods which maximize the sample size using imputation would result in better performance. Due to the non linearity and complexity of multi modal biological data, we decided to focus on deep learning techniques.  

\begin{figure}[tb]
\begin{center}
\includegraphics[width=0.98\linewidth]{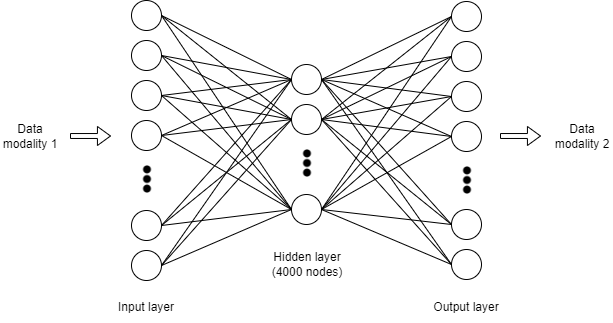}
\caption{Autoencoder based architecture used for deep imputation of modalities.}
\label{fig:tdimpute_architecture}
\end{center}
\end{figure}

We have extended the work described by \citet{zhou2020imputing}, which only imputes RNAseq values from methylation data, to more than two data modalities and demonstrated this approach on  RNA-to-methylation and miRNA-to-RNA imputations. For practical reasons, patient data from clinical trials or real world evidence are often incomplete and only a small portion of samples feature all multi-omics measurements. Importantly, it is common to measure only a very limited number of features within a modality (e.g. a gene expression panel may include only about 1,000 genes out of 20,000). Therefore, using a model trained on a cohort of full multi-omics measurements, missing data in another sample set can be imputed using its existing modality only, resulting in a complete  multi-omics data set which can be used in downstream prediction tasks. The neural network shown in Figure \ref{fig:tdimpute_architecture} was used for imputation of whole missing feature vectors. The network uses architecture parameters from the original paper and is a three layer network with a bottleneck containing 4,000 nodes. This network is fully connected and uses a sigmoid activation function trained with the ADAM optimiser \citep{kingma2014adam}. No dropout was applied as it was shown to decrease performance. The loss function optimised during training was the root mean squared error (RMSE) between the original and imputed values. 

\subsection*{Multitask learning}\label{multitask_methods}

\citet{zhang2021omiembed} analysed three modalities, RNAseq, miRNA and methylation in combination and individually to investigate the benefit of using multiple modalities for the prediction of three independent target variables in a multitask setting. However, in this experimental setting, only patients with data for all three modalities and all three tasks could be used in the experiments. Patients with one or more missing modalities were excluded, and any missing values within an individual modality were imputed using mean imputation.

\begin{figure*}[tb]
\begin{center}
\centering
\includegraphics[width=\linewidth]{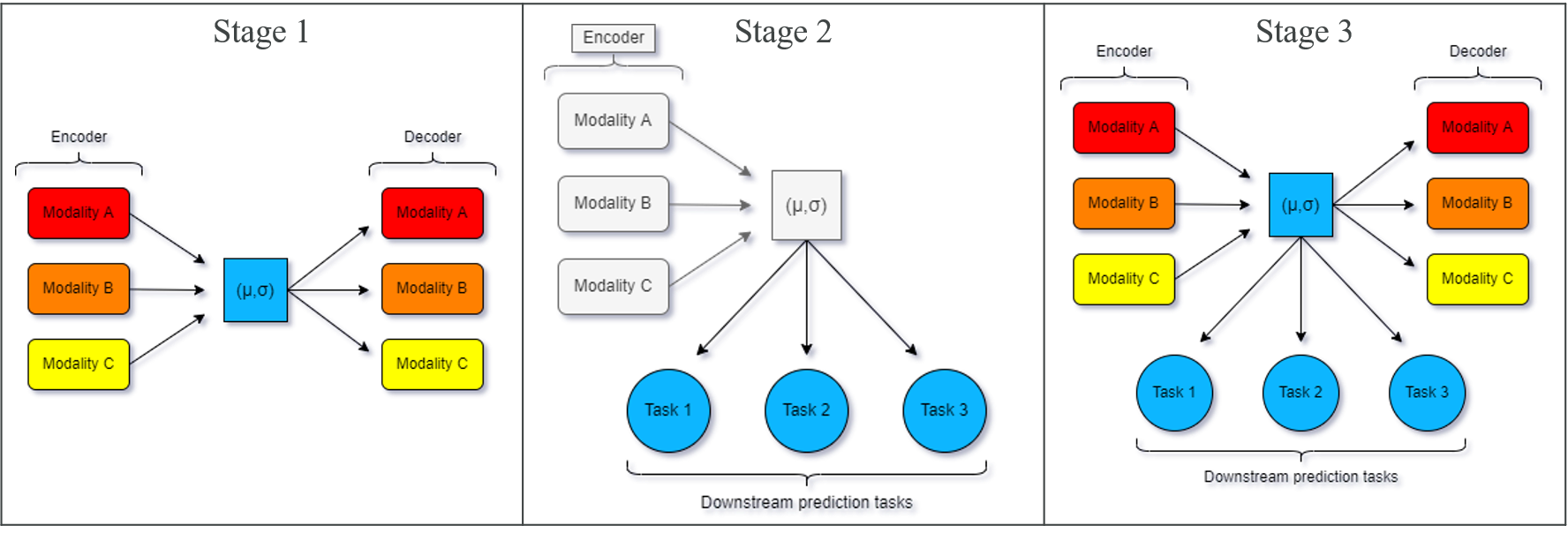}
\caption{Combining data modalities in multitask learning architecture. An embedding module generates a low-dimensional representation of the data, then a downstream task module calculates predictions for three related tasks. Training consists of three distinct stages. Left: in stage 1 the deep embedding module is trained; middle: in stage 2, weights in the deep embedding module are frozen and training is performed for the downstream tasks; right: in stage 3 all weights in the network are fine-tuned.}
\label{fig:omiembed_architecture}
\end{center}
\end{figure*}

Here, the model we present is composed of two modules as shown in Figure \ref{fig:omiembed_architecture} - a deep embedding module followed by a downstream task module. Since omics data types are typically high dimensional \citep{bellman1961}, a regularised, low-dimensional representation of the data was obtained using a deep embedding module. All modalities were concatenated together before being passed as input to the deep embedding module that is based on a variational autoencoder (VAE) \citep{kingma2013auto}. This network learnt a joint embedding mapping from the multiple data types to a low-dimensional latent space. The encoded vector was fed through two bottleneck layers to obtain mean and standard deviation vectors which define the Gaussian distribution and in turn the variational distribution. The latent variable was then fed into the decoder part of the VAE to reconstruct the input and train the network.

The latent representation was also the input to the downstream task module. This module is formed of three networks, each of which deals with one task. The types of task available to the user are diagnostic (classification), prognostic (survival) and demographic (regression), and the model can train either one task alone, or all three together to exploit shared information which may be helpful in more than one of these related tasks. The motivation is that sharing representations between related tasks can improve generalisation \citep{caruana1997multitask}. For classification and regression tasks, fully connected multi-layer networks were used, and for survival analysis an adapted version of multitask logistic regression was employed with the dimension of the output layer being the number of time intervals the time axis was divided into \citep{yu2011learning}. Dropout layers were included in these downstream networks to prevent overfitting \citep{srivastava2014dropout}.

Training was performed in three stages. The first stage was unsupervised and sought to optimise the deep embedding parameters of the model. Weights in the downstream task networks were fixed, and the loss function in Equation \ref{eq:embed_loss} was minimised to update the weights in the deep embedding module. Here, $\mathbf{x}_j$ and $\mathbf{x}_j^{'}$ denote the input and reconstructed data vectors, where $j$ is the index of the current modality and $M$ is the total number of modalities. The first term calculates the binary cross entropy between the input and reconstructed data, and the second term is the KL divergence between the learned distribution for the latent dimension and a unit Gaussian. Experiments were carried out to determine that the two first stages where sub-parts of the model are trained separately are necessary for the best possible survival predictions.

\begin{equation}
  \label{eq:embed_loss}
    L_{embed} = \frac{1}{M} \sum_{j=1}^{M} BCE ( \mathbf{x}_j,\mathbf{x}_j^{'}) + D_{KL} (\mathcal{N} (\bm{\mu},\bm{\sigma}) || \mathcal{N} (\mathbf{0}, \mathbf{I}))
\end{equation}

In the second stage of training, the weights in the deep embedding module were fixed and the downstream networks trained simultaneously in a supervised fashion. As seen in Equation \ref{eq:down_loss}, a joint loss function was minimised for all downstream tasks, which is the sum of the individual loss functions where the relative weights of each component can be adjusted using the parameter $w_k$. Here $K$ is the total number of downstream networks.

\begin{equation}
  \label{eq:down_loss}
    L_{down} = \frac{1}{K} \sum_{k=1}^{K} w_k L_{down_k}    
\end{equation}

In the third training stage the loss functions from each part of the model were combined in a sum weighted by $\lambda$, giving the loss function shown in Equation \ref{eq:total_loss}. In this final stage of training all weights in the model were fine-tuned across the deep embedding and downstream task networks to obtain a better performance. 

\begin{equation}
  \label{eq:total_loss}
  L_{total} = \lambda L_{embed} + L_{down}
\end{equation}

\subsection*{Multitask learning with deep imputed data}

Following on from these two distinct pieces of work, we extended and generalised the deep imputation method outlined in \citet{zhou2020imputing} to handle several modalities with missing data. This enabled us to compensate for the cases where some of the patients have less than the three modalities as encountered by \citet{zhang2021omiembed}. On top of imputing RNAseq data from DNA methylation data, we extended the work to also impute DNA methylation data using RNAseq data, and impute miRNA data using RNAseq data. In this way, we were able to predict values for patients who had data for two modalities but were missing the third one, resulting in a significant increase of 19\% in the number of patients used in the multi-task learning framework. In the next section we demonstrate how our methodology applied to a pan-cancer TCGA dataset led to promising results. 

\section{Experiments}\label{experiments}

Experiments were carried out using multi-omics pan-cancer TCGA data \citep{tcga}. This consisted of RNAseq, methylation and miRNA data from patients across 33 cancer types. Experiments were executed on two Volta GPUs with 96GB of memory each. Time taken to run the experiments ranged from 10 minutes to 1 hour depending on number of modalities used, dimensionality of each modality and number of epochs in each training stage. Supporting code is available on github \footnote{\href{https://github.com/sophiep96/multitask_impute}{https://github.com/sophiep96/multitask\_impute}}.

Through a series of experiments we compared the effect of using the original data compared to deep imputed data. These two datasets were used in three tasks - a classification task predicting cancer type, a regression task for age, and survival prediction. The results were compared using separate models for each task, versus a multitask model in which all three tasks are trained together. We summarise the results in four parts:

\begin{enumerate}[label=(\Alph*)]
    \item Using the original data without deep imputation and training tasks separately;
    \item Using the original data without deep imputation and multitask learning;
    \item Using deep imputed data and training tasks separately;
    \item Using deep imputed data and multitask learning.
\end{enumerate}

In each subsection we compare results achieved using different combinations of the three available data modalities. 

\subsection*{Setup for experiments A and B (without deep imputation)}

The dimensionality of the different modalities varies drastically - the raw RNAseq data has $\sim$20,000 genes, methylation data has $\sim$450,000 probes and miRNA has $\sim$1,800 genes. All RNAseq genes without zero values for every patient were retained and a $log_2$ transformation applied. For DNA methylation data, sites with more than 90\% missing data and those situated on the Y chromosome were excluded. Then, to decrease the computational load during training, 50\% of the probes were randomly chosen, resulting in $\sim$200,000 probes. For miRNA data, the dimensionality is relatively low so no dimensionality reduction was required. A $log_2$ transformation was applied to miRNA data. All missing values within each modality were imputed using mean imputation, and then scaled to $[0,1]$ range using MinMaxScaler from scikit-learn \citep{pedregosa2011scikit} for better convergence of the deep learning networks.

\subsection*{Setup for experiments C and D (using deep imputed data)}

For deep imputation, the same preprocessing steps were followed except a lower dimensionality was used for the RNAseq and methylation data. This step was taken to mitigate the huge difference in dimensionality between the three modalities - miRNA had a dimensionality of only roughly 1,800 while RNAseq and DNA methylation had 20,000 and 450,000 respectively. The 5,000 most variable dimensions were therefore retained for both RNAseq and DNA methylation data since this demonstrated faster convergence and better results during training for the deep imputation stage of the pipeline. 

In the setup for experiments A and B, patients were only included if they had data for all three modalities. In order to maximise the size of the dataset and therefore the predictive power in experiments C and D, we used deep imputation to allow inclusion of patients who were missing data from one modality. The work of \citep{zhou2020imputing} was extended and generalised to be able to impute more modalities than just DNA methylation. We were therefore able to impute missing RNAseq data from DNA, DNA data from RNAseq, and miRNA from RNAseq. RNAseq and DNA datasets were chosen as inputs for training due to the poor performance of miRNA data alone in predictive tasks. RNAseq was chosen to impute miRNA since we used the 5,000 most variable dimensions due to computational constraints. The original dimensionality of RNAseq was much lower than that of methylation data, so we reasoned that the top 5,000 genes would be relatively more informative. Table \ref{tab:imputed_data} shows the breakdown of the overall effect of this deep imputation step on the complete data set. Data for a total of 1,542 patients was imputed, increasing the size of the cohort from 8,129 to 9,671 patients. The majority (88\%) of the patients who we imputed data for were missing DNA methylation data. 

\begin{table}
    \centering
    \caption{No. patients with imputed data for one modality.}
    \begin{tabular}{ccc} \hline
        Missing modality & Modality used to impute & No. of patients \\ \hline
        RNA & Methylation & 371\\
        Methylation & RNA & 1,358\\
        miRNA & RNA & 173\\ \hline
         & & 1,542\\\hline
    \end{tabular}
    \label{tab:imputed_data}
\end{table}

\section{Results}\label{results}

\subsection{Training tasks separately with original data}\label{results_a}

This set of experiments used the original dataset without deep imputation, so only patients with data for all three modalities were included (a total of 8,129 patients). Separate models were trained for prediction of cancer type (classification task), age (regression task) and survival. Experiments were carried out for each modality separately, as well as using the combination of RNAseq with DNA methylation, and the combination of all three modalities. The results for these experiments can be seen in `Experiment A' in Table \ref{tab:results}, which shows the mean scores for each metric over 5 cross validation folds, and the standard deviation across these folds.

In the case of single task models for classification and regression, the best performance was achieved using all modalities. The classification model using all modalities achieved an average AUC (area under the ROC curve) of 0.9997 and F1 of 0.965 across five cross validation folds, while the regression model using all modalities had an average RMSE of 10.252 and $R^2$ of 0.496. For the survival single task model, the best performance was achieved using just DNA (c-index 0.739) and the model using all available modalities resulted in the second best performance (c-index 0.729). This shows that for single tasks trained in isolation, combining modalities is beneficial. Figure \ref{fig:training_plots} in the Appendix shows examples of metric scores through the training process for classification, regression and survival predictions. Additional information on the training stages is included also in the Appendix.

\subsection{Multitask learning with original data}\label{results_b}

The same data was used in this experiment as in the previous section. However, in this setting all three tasks were trained simultaneously using the architecture outlined by \citet{zhang2021omiembed}, with the diagnostic, demographic and prognostic tasks being prediction of cancer type, age and survival respectively. Experiments were done using the same combinations of modalities as above and results are shown in `Experiment B' of Table \ref{tab:results}. For this set of experiments, the best performance for both the classification and regression tasks was achieved using RNA data alone. The results for these models were AUC 0.999, F1 0.943 for classification and RMSE 11.891, $R^2$ 0.323 for regression. For the survival task, the best performance of c-index 0.751 is achieved using all available modalities, closely followed by DNA alone (c-index 0.750).

\subsection{Training tasks separately with deep imputed data}\label{results_c}

This set of experiments used the extended dataset generated through deep imputation. Patients with missing data for one modality could therefore be included, increasing the total dataset size to 9,671 patients. `Experiment C' in Table \ref{tab:results} contains the results from experiments carried out for single task training with the larger dataset including imputed data for patients who initially only had two out of three modalities available. 

The best performance for all tasks in this setting was achieved using multiple modalities. For classification the best result of mean AUC 0.999 and F1 0.960 was achieved using all three modalities but performance was slightly lower than without the imputed data. For regression and survival the best results were achieved using a combination of just RNAseq and methylation data (for regression this was mean RMSE 9.994 and $R^2$ 0.506, and for survival was c-index 0.721). Both models yielded performance better than the corresponding one achieved without imputed data (RMSE 11.548, $R^2$ 0.361, c-index 0.703). Table \ref{fig:training_plots} in Appendix shows examples of metric scores through the training process for classification, regression and survival predictions as well as more information on the training stages.

\subsection{Multitask learning with deep imputed data}\label{results_d}

Here, as in the previous section, deep imputed data was used with the same modality combinations as above. All three tasks were trained simultaneously using the architecture outlined by \citet{zhang2021omiembed}, with the diagnostic, demographic and prognostic tasks being prediction of cancer type, age and survival. For Experiment D, Table \ref{tab:results} shows that RNAseq alone leads to the best performance for the classification task (AUC 0.999, F1 0.915). However, using multiple modalities is useful for both of the other tasks; the best regression performance is achieved using all three modalities (RMSE 11.628, $R^2$ 0.331), and the best survival performance of c-index 0.735 is achieved using the combination of RNA and DNA modalities. 

\begin{table*}
\caption{Results for four sets of experiments (A, B, C, D) with each combination of data modalities. Bold numbers indicate which combination of modalities led to the best results for each downstream task with a single set of experiments. \textcolor{Green}{\underline{Green underlined values in Experiments B and C are better than the corresponding results in Experiment A.}}}
\begin{center}
\begin{tabular}{@{}lcrrrrr@{}}
\specialrule{.2em}{.1em}{.1em}
&&\multicolumn{5}{c}{Modality}\\
\cmidrule(l){3-7}
Task&Metric&\multicolumn{1}{c}{RNA}&\multicolumn{1}{c}{DNA}&\multicolumn{1}{c}{miRNA}&\multicolumn{1}{c}{RNA+DNA}&\multicolumn{1}{c}{All}\\
\specialrule{.2em}{.1em}{.1em}
\addlinespace[2pt]
\multicolumn{7}{c}{\bfseries Experiment A: training tasks separately with original data.}\\
\specialrule{.2em}{.1em}{.1em}
\addlinespace[4pt]
\multirow{2}{*}{Classification}& F1 & 0.945 $\pm$ 0.003 & 0.941 $\pm$ 0.004 & 0.263 $\pm$ 0.011 & 0.905 $\pm$ 0.018 & \textbf{0.965} $\pm$ \textbf{0.004}\\ 
& AUC & 0.999 $\pm$ 0.001 & 0.997 $\pm$ 0.002 & 0.841 $\pm$ 0.006 & 0.998 $\pm$ 0.001 & \textbf{0.9997} $\pm$ \textbf{0.000} \\ 
\midrule
\addlinespace[4pt]
\multirow{2}{*}{Regression} & RMSE & 10.745 $\pm$ 0.232 & 10.726 $\pm$ 0.704 & 14.358 $\pm$ 0.252 & 11.548 $\pm$ 0.175 & \textbf{10.252} $\pm$ \textbf{0.467} \\ 
& $R^{2}$ & 0.447 $\pm$ 0.010 & 0.446 $\pm$ 0.074 & 0.012 $\pm$ 0.011 & 0.361 $\pm$ 0.021 & \textbf{0.496} $\pm$ \textbf{0.140}\\ 
\midrule
\addlinespace[4pt]
Survival & c-index & 0.714 $\pm$ 0.019 & \textbf{0.739} $\pm$ \textbf{0.021} & 0.529 $\pm$ 0.008 & 0.703 $\pm$ 0.018 & 0.729 $\pm$ 0.027 \\ 
\specialrule{.2em}{.1em}{.1em}
\addlinespace[2pt]
\multicolumn{7}{c}{\begin{minipage}[t]{0.7\textwidth}\bfseries\centering Experiment B: multitask learning with original data.\end{minipage}}\\
\specialrule{.2em}{.1em}{.1em}
\addlinespace[4pt]
\multirow{2}{*}{Classification}  & F1 & \textbf{0.943} $\pm$ \textbf{0.002} & 0.594 $\pm$ 0.042 & 0.216 $\pm$ 0.010 & 0.678 $\pm$ 0.027 & 0.822 $\pm$ 0.016\\ 
& AUC & \textbf{0.999} $\pm$ \textbf{0.001} & 0.993 $\pm$ 0.001 & 0.883 $\pm$ 0.004 & 0.998 $\pm$ 0.001 & 0.999 $\pm$ 0.000 \\ 
\midrule
\addlinespace[4pt]
\multirow{2}{*}{Regression}& RMSE & \textbf{11.891} $\pm$ \textbf{0.290} & 12.348 $\pm$ 0.144 & 14.365 $\pm$ 0.270 & 12.385 $\pm$ 0.290 & 12.370 $\pm$ 0.612\\ 
& $R^{2}$ & \textbf{0.323} $\pm$ \textbf{0.016} & 0.270 $\pm$ 0.016 & 0.012 $\pm$ 0.016 & 0.266 $\pm$ 0.027 & 0.264 $\pm$ 0.080\\ 
\midrule
\addlinespace[4pt]
Survival & c-index & \textcolor{Green}{\underline{$0.732 \pm 0.006$}} & \textcolor{Green}{\underline{$0.750\pm 0.010$}} & \textcolor{Green}{\underline{$0.542\pm 0.020$}}& \textcolor{Green}{\underline{$0.745\pm 0.014$}} & \textcolor{Green}{\underline{$\textbf{0.751}\pm \textbf{0.017}$}}\\ 
\specialrule{.2em}{.1em}{.1em}
\addlinespace[2pt]
\multicolumn{7}{c}{\begin{minipage}[t]{0.7\textwidth}\bfseries\centering Experiment C: training tasks separately with deep imputed data.\end{minipage}}\\
\specialrule{.2em}{.1em}{.1em}
\addlinespace[4pt]
\multirow{2}{*}{Classification} & F1 & 0.939 $\pm$ 0.006 & 0.922 $\pm$ 0.013 & \textcolor{Green}{\underline{$0.920\pm0.003$}} & \textcolor{Green}{\underline{$0.952\pm0.009$}} & \textbf{0.960} $\pm$ \textbf{0.006}\\ 
& AUC & 0.999 $\pm$ 0.000 & \textcolor{Green}{\underline{$0.998\pm 0.001$}} & 0.998 $\pm$ 0.000 & 0.999 $\pm$ 0.000 & \textbf{0.999} $\pm$ \textbf{0.001} \\ 
\midrule
\addlinespace[4pt]
\multirow{2}{*}{Regression}& RMSE & \textcolor{Green}{\underline{$10.683\pm 0.073$}} & \textcolor{Green}{\underline{$10.581\pm 0.092$}} & \textcolor{Green}{\underline{$12.227\pm 0.151$}} & \textcolor{Green}{\underline{$\textbf{9.994}\pm \textbf{0.086}$}} & \textcolor{Green}{\underline{$10.196\pm 0.121$}} \\ 
 
& $R^{2}$ & 0.443 $\pm$ 0.020 & 0.446 $\pm$ 0.008 & \textcolor{Green}{\underline{$0.260\pm 0.008$}} & \textcolor{Green}{\underline{$\textbf{0.506}\pm \textbf{0.012}$}} & 0.486 $\pm$ 0.013\\ 
\midrule
\addlinespace[4pt]
Survival & c-index & \textcolor{Green}{\underline{$0.720\pm 0.017$}} & 0.657 $\pm$ 0.032 & \textcolor{Green}{\underline{$0.672\pm 0.017$}} & \textcolor{Green}{\underline{$\textbf{0.721}\pm \textbf{0.018}$}} & 0.719 $\pm$ 0.020 \\ 
\specialrule{.2em}{.1em}{.1em}
\addlinespace[2pt]
\multicolumn{7}{c}{\begin{minipage}[t]{0.7\textwidth} \bfseries\centering Experiment D: multitask learning with deep imputed data. \\The combination of deep imputation and multitask learning did not result in better performance compared to Experiments B and C.\end{minipage}}\\
\specialrule{.2em}{.1em}{.1em}
\addlinespace[4pt]
\multirow{2}{*}{Classification}  & F1 & \textbf{0.915} $\pm$ \textbf{0.006} & 0.853 $\pm$ 0.005 & 0.787 $\pm$ 0.023 & 0.897 $\pm$ 0.010 & 0.903 $\pm$ 0.008\\ 
& AUC & \textbf{0.999} $\pm$ \textbf{0.001} & 0.996 $\pm$ 0.001 & 0.995 $\pm$ 0.001 & 0.998 $\pm$ 0.001 & 0.999 $\pm$ 0.000 \\ 
\midrule
\addlinespace[4pt]
\multirow{2}{*}{Regression}& RMSE & 12.092 $\pm$ 0.077 & 11.924 $\pm$ 0.068 & 12.578 $\pm$ 0.105 & 11.926 $\pm$ 0.102 & \textbf{11.628} $\pm$ \textbf{0.124}\\ 
& $R^{2}$ & 0.277 $\pm$ 0.008 & 0.297 $\pm$ 0.006 & 0.217 $\pm$ 0.016 & 0.297 $\pm$ 0.006 & \textbf{0.331} $\pm$ \textbf{0.011}\\ 
\midrule
\addlinespace[4pt]
Survival & c-index & 0.729 $\pm$ 0.033 & 0.704 $\pm$ 0.026 & 0.695 $\pm$ 0.039 & \textbf{0.735} $\pm$ \textbf{0.021} & 0.726 $\pm$ 0.020\\ 
\specialrule{.2em}{.1em}{.1em}
\end{tabular}
\end{center}
\label{tab:results}
\end{table*}

\begin{figure*}
\begin{center}
\hspace*{-1in}
\centering
\includegraphics[width=1.25\linewidth]{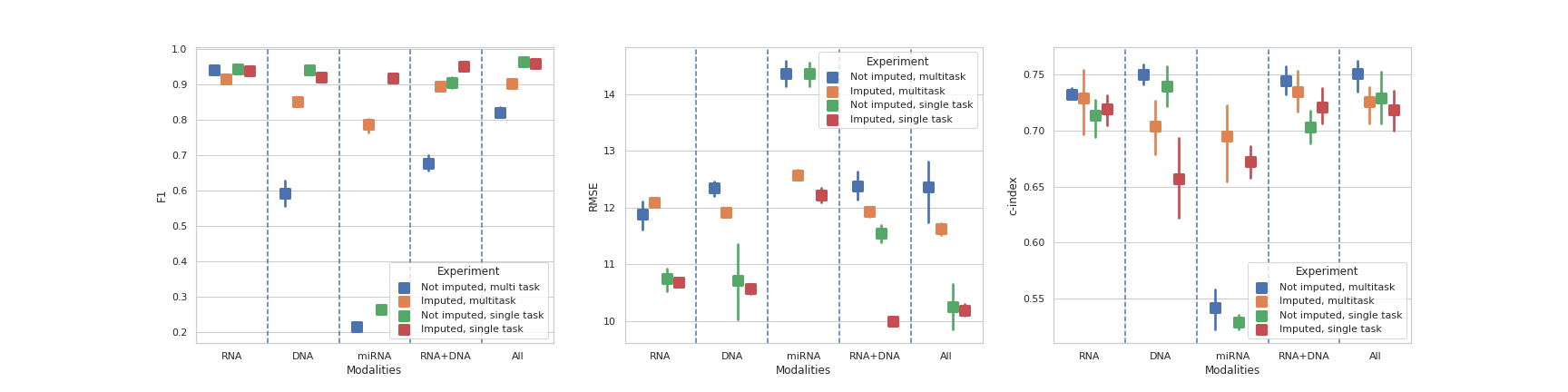}
\caption{Performance of single task vs. multitask and non-imputed vs. deep imputed data for each downstream task, across each combination of modalities tested. Left: classification models - best performance is achieved using single task models, with both non-imputed and deep imputed data; middle: regression models - deep imputation improves performance and model stability across cross-validation folds; right: survival models - multitask learning leads to better performance across all modalities.}
\label{fig:multitask_scatter}
\end{center}
\end{figure*}

\subsection*{Multitask learning removing training stages}

The best performance for survival prediction was achieved in Experiment B using multitask learning. Further experimentation was carried out in this setting to determine whether all three training stages were beneficial for the prediction. We did three experiments (removing stage 1, removing stage 2, and removing both stages 1 and 2) and the total number of epochs in all experiments was kept constant. These models achieved statistically significantly worse survival predictions in each of these settings compared to using all three training stages. These results confirmed that all three stages of the multitask training framework are beneficial for survival predictions.

\section{Discussion}

Overall our results are comparable to those of \citet{zhang2021omiembed}, although differences in data preprocessing means we do not expect them to be exactly the same. In the case of  survival analysis, the results of \citet{zhang2021omiembed} are  better (c-index 0.751 vs 0.782), while for classification the results are comparable (F1 0.965 vs 0.968). Lastly for regression, our results demonstrated better performance (RMSE 9.994 vs 10.634). It was not possible for us to calculate statistical significance of the difference between results since the performance of individual cross validation splits was not available for the other paper.

\subsection*{Classification models}

In the left panel of Figure \ref{fig:multitask_scatter} we observe that the four experiments for classification had similar performance when the input modality was RNA expression alone. With regards to miRNA, deep imputation led to significantly better performance compared to the results without imputed data. This is expected since 1,542 new patients were added to the dataset but only 173 of them had imputed miRNA data. The number of patients with original miRNA data therefore increased significantly (these patients had either RNA or DNA methylation data imputed). From the miRNA results without imputation it seems that there was not enough useful information present to successfully predict the chosen tasks. The imputation step also introduced information from the higher performing RNAseq data into the miRNA dataset. miRNA data alone is not able to achieve good performance, but adding in the extra information from RNAseq data boosts performance significantly.

For the other three modality combinations, non-imputed data with multitask learning performs worse than the rest of the models. For all classification models the error bars are very small, showing that the models are stable across cross validation folds. Across all experiments the best performance was achieved using a single task model with all modalities and no deep imputation. However cancer type classification by itself proved to be an easy task - most models had $AUC>0.99$ and $F1>0.90$. $F1$ scores were lower in the multitask setting but were vastly improved by using imputed data. Our key finding for the classification task is that the best performance is achieved using single task models, although whether this model performs better with the original or deep imputed data varies across modalities. For this application, deep imputation was partially useful in improving predictive power.

\subsection*{Regression models}

The centre panel of Figure \ref{fig:multitask_scatter} shows the performance of regression models measured by RMSE (so a lower score is better). Here, the error bars are larger than for classification, particularly for models using non-imputed data; increasing the dataset size using deep imputation has improved the stability across cross validation folds. The figure shows that while imputation improves performance for miRNA alone, the best performance is roughly the same as for the other modalities with non-imputed data. This plot shows that training single task models results in the best performance for regression models. However this does not mean that the information learned in the multitask models is not improving the predictions of survival. Since survival analysis is the most clinically relevant task, improving the survival predictions at the expense of regression performance can still be viewed as a success. Our key finding in regression models is that single task models perform best, using RMSE for performance evaluation. 

Across all modality combinations deep imputation slightly improved the performance of models and significantly improved stability across cross validation folds. The best regression performance was achieved by a single task model using RNAseq and methylation data with deep imputation. It is encouraging that better results were achieved using deep imputed methylation data in both single and multitask settings (compared to the corresponding experiments without imputed data). The majority of new patients had imputed data for methylation, showing that the information captured in imputed values is meaningful in predicting these endpoints.

\subsection*{Survival models}

The right panel of Figure \ref{fig:multitask_scatter} again shows the large improvement for miRNA alone, where the use of deep imputed data improves performance. The opposite can be seen for DNA methylation. Since both modalities were imputed using RNAseq data, this suggests that information in the original miRNA data is orthogonal to that in RNAseq data. Therefore combining the two modalities by using RNAseq for deep imputation of miRNA leads to additional information relevant to survival being encoded in the miRNA data. Unexpectedly, no extra information gain was observed when imputing DNA methylation from RNAseq data. 

For all modality combinations apart from miRNA alone, the best performing survival model is a multitask model trained with non-imputed data. The error bars here are fairly large and are consistent across experiments. This was expected of survival models since we had to use a higher learning rate and fewer epochs to avoid overfitting, resulting in less stable models. For both non-imputed and deep imputed data, multitask training gains an improvement for survival prediction, although at the expense of performance in the other two tasks. Even though the performance for classification and regression decreases, the information required for these tasks is beneficial for the survival task and therefore leads to an improvement in c-index. This exemplifies the key idea motivating multitask learning - the information required for related tasks is also beneficial for this task and reduces the redundancy in feature learning which happens when training tasks individually. The best survival performance is achieved by the model using all three modalities in a multitask setting with no deep imputation (c-index 0.751). For all combinations of modalities except for just RNAseq, the classification and regression performance in the multitask setting (Experiment D) was an improvement on using the original data (Experiment B). 

Our key finding for survival models is that the best performance was obtained by using multitask learning and no imputation when using all modalities together. Combining RNAseq and DNA data led to model performance which considerably overlaps with that of the model using multitask learning and deep imputation. Multitask learning led to a significant improvement in survival prediction across modalities.

\section{Conclusions and Future Work}\label{conclusion}

In this paper, we investigated how deep imputation in combination with multi-modality and multitask learning can be used to reach state-of-the-art prediction results using combined -omics modalities (RNA, DNA methylation and miRNA). This  was demonstrated to be a promising approach where some patients have missing data for a modality. We extended and generalised a previous deep imputation method to imputation of RNA from DNA methylation as well as miRNA from RNA and demonstrated its usefulness in experiments on TCGA data. The generalised approach allowed us to impute the missing values and achieve better downstream performance. Moreover, adding multitask learning was shown to be very useful in extracting information from high dimensional multi-omics data, superior to solving a single task. Given the tasks of predicting cancer type, age and survival, we saw that multitask learning was particularly useful for survival analysis, and deep imputation was useful in regression models. Deep imputation alone outperformed multitask learning in classification and regression tasks across all combinations of modalities, while multitask learning alone outperformed deep imputation when combining several modalities for survival prediction. We conclude that both approaches are valuable when optimising performance for downstream predictive tasks. We finish with some key learnings from this work:

\begin{enumerate}
    \item Deep imputation allowed us to increase the number of samples in the dataset by 19\%, leading to significant increases in performance, especially for models using miRNA data.
    \item Deep imputation helped significantly more for some modalities (e.g. miRNA by itself, and RNA+DNA) than for others (e.g. RNA by itself). Users should therefore carry out experiments to determine the benefit of imputation on their modalities.
    \item Training multitask models resulted in better survival results than training a single task survival model. This is true for every combination of modalities, using both non-imputed and deep imputed data.
    \item Even though both deep imputation and multitask learning boosted performance on their own, combining them together did not bring any further improvements.
    \item Using multiple modalities is nearly always beneficial, at least for the modalities of miRNA, RNA and methylation.
\end{enumerate}

Further work could be done looking into incorporating other modalities (e.g. embeddings from digital pathology images) and other datasets. Although TCGA is a large and well characterised dataset composed of many different cancer types, it would still be very beneficial to look into how well the approaches in this paper work in other datasets to fully test the performance of the model. Potentially further hyper-parameter tuning could find multitask learning strategies which achieve even better results, in particular which are more stable during training for survival models. As well as altering the way the current models are trained, further work could be done using different deep learning architectures to do both the deep imputation and multitask learning stages of the pipeline. 

\section{Acknowledgment}\label{acknowledgments}
Deepest appreciation to AstraZeneca, UK for hosting the research project and to Nikos Nikolaou and Jan Zaucha who helped to review the manuscript.

\bibliographystyle{IEEEtranN}
\bibliography{imputation_bib} 
\newpage
\section{Appendix}\label{appendix}

\begin{table*}
\begin{center}
\vspace*{-1in}
\begin{tabular}{m{1.5cm}m{6cm}m{6cm}}
& \Centering Original data & \Centering Deep imputed data \\
Classification & \subfloat[AUC - most learning happens at the start of stage 3 of training]{\includegraphics[width=6cm]{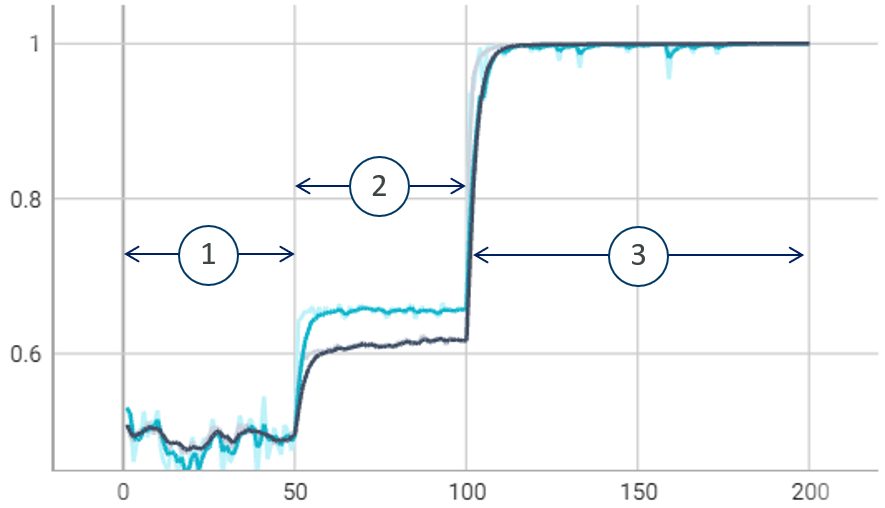}} & \subfloat[AUC in imputed model - most learning happens at the start of stage 2 of training]{\includegraphics[width=6cm]{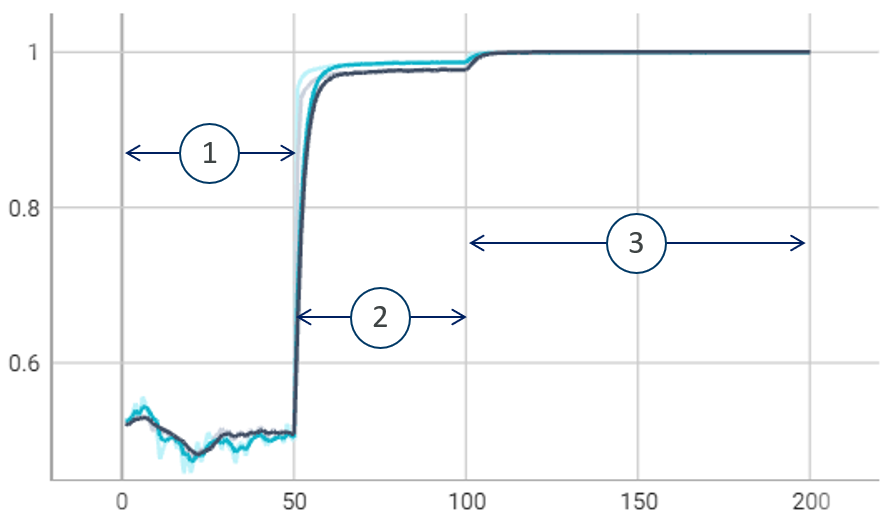}} \\
Regression & \subfloat[RMSE - most learning happens at the start of stage 2 of training]{\includegraphics[width=6cm]{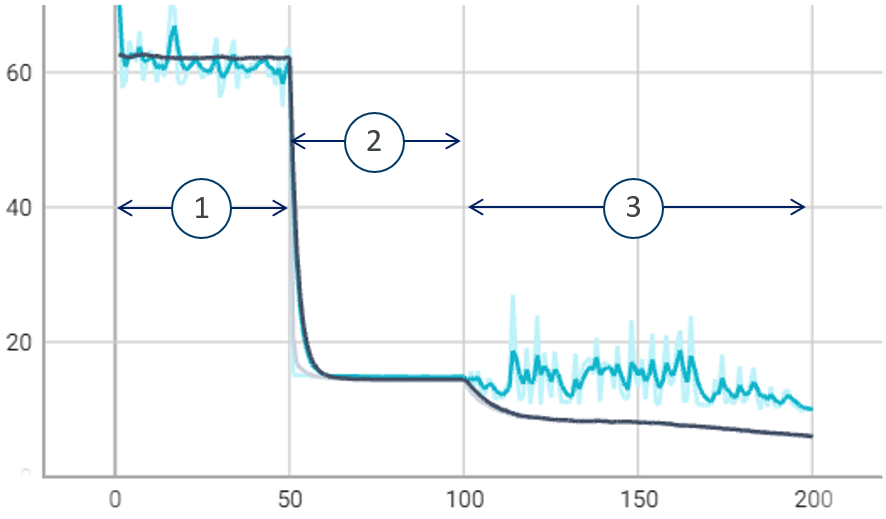}} & \subfloat[RMSE in imputed model - most learning happens at the start of stage 2 of training]{\includegraphics[width=6cm]{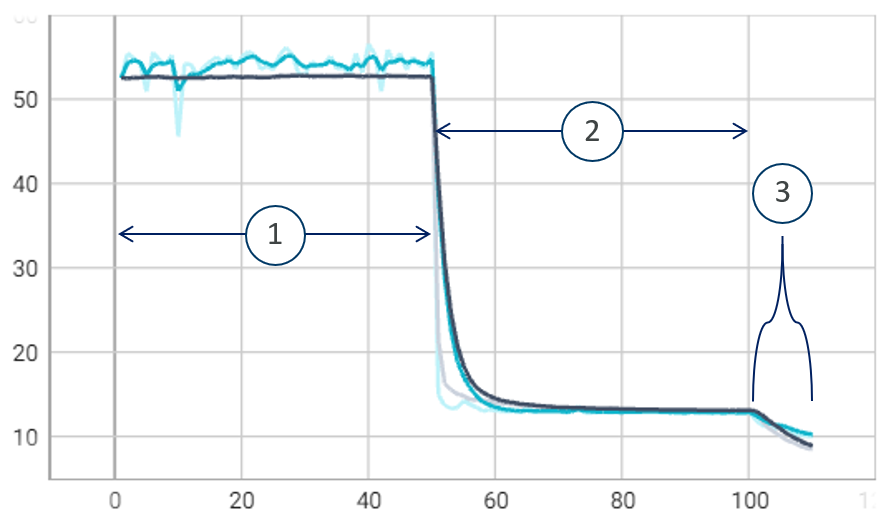}} \\
Survival & \subfloat[c-index - most learning happens in stage 3 of training, and training is stopped early to prevent overfitting]{\includegraphics[width=6cm]{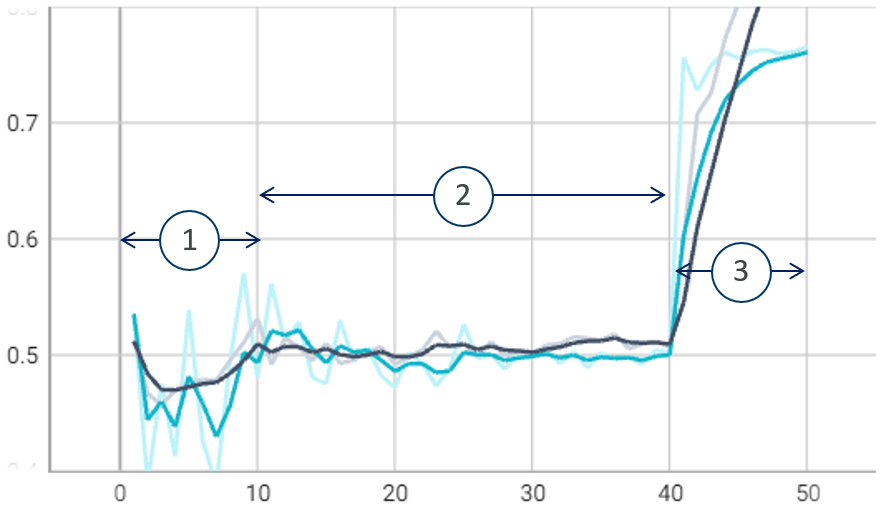}} & \subfloat[c-index in imputed model - most learning happens in stage 3 of training, and training is stopped early to prevent overfitting]{\includegraphics[width=6cm]{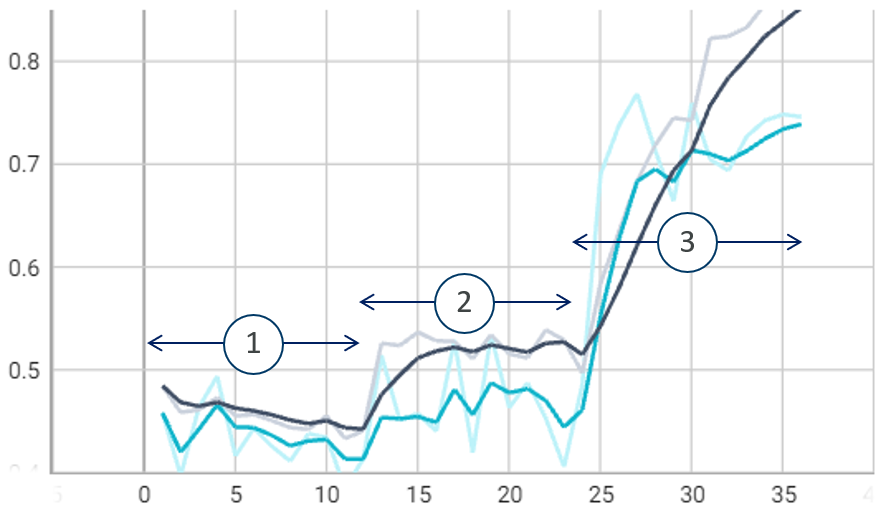}} \\
\end{tabular}
\caption{The change in AUC, RMSE, c-index ($y$-axis) through training of models for a number of epochs ($x$-axis). The black line is the score on the training set and the blue line is the score on the test set with a smoothing factor of 0.6 applied. Model performance during stage 1 is a baseline, as the encoder is being trained, but the latent space representations are then passed into downstream networks with random weights (no training is done on the downstream network arms during stage 1).}
\label{fig:training_plots}
\end{center}
\end{table*}

\subsection{Training stages and the process without imputation}
 The training plots in the first column of Figure \ref{fig:training_plots} are for single task models using all three modalities together with original data. As detailed in Section \ref{multitask_methods}, there are three stages of training in which the embedding module is first trained, then the downstream networks, and finally both parts are fine-tuned together. The three stages can be identified by sharp improvements in performance. In the classification model predicting cancer type, the third stage of training with fine-tuning of the whole model results in the most pronounced improvement in AUC. In the regression model predicting age the converse is true. We do not see a large improvement in stage 3, but see a large improvement at the start of stage 2 when the weights of the downstream network are updated. There is a slight further improvement during the fine-tuning stage. Both the classification and regression tasks were trained for a total of 200 epochs across all 3 training stages using a learning rate of 0.001. However, training for this long resulted in worse results for the survival model. The best results were gained using a lower learning rate and training for fewer epochs (a total of 50 across the 3 training stages with a learning rate of 0.0003). In the figure it still looks like the model is improving at the point where training is stopped. However, we found that if we trained for any longer the model was overfitting and test set performance decreased. 
 
 \subsection{Training stages and the process with imputation}
 The training plots in the second column of Figure \ref{fig:training_plots} are for single task models using all three modalities with deep imputed data. The classification model was trained for 200 epochs with a learning rate of 0.0001. Here we can see a steep increase in performance at the 50 epoch mark at the start of stage 2. The third stage of training where fine-tuning takes place did not have much effect for this model. The same is true for the regression model, which was trained for 110 epochs with a learning rate of 0.0003. There is a sharp improvement at the start of stage 2 but not stage 3. For this model a smaller number of epochs (50 in each of the first two stages and 10 in stage 3) resulted in the best performance. The training curves for the survival model are less consistent within each stage. However there is a sharp increase in performance at the start of stage 3. Like in the survival model with non-imputed data, we found that training for a longer time quickly led to overfitting, hence why the model was trained for fewer epochs (36 epochs in total) with a higher learning rate (0.001) and training was stopped at a time when it looks like the model could improve further. In fact we found that if training continues the model overfits and performance on the test set decreases.

\end{document}